\input epsf
\documentstyle[twocolumn,floats,aps,epsf]{revtex}
\begin{document}
\twocolumn[\hsize\textwidth\columnwidth\hsize\csname
@twocolumnfalse\endcsname

\title{Temperature Dependent Resistivity of Single Wall Carbon Nanotubes}

\author{C.L. Kane$^1$, E.J. Mele$^1$, R.S. Lee$^2$, J.E. Fischer$^2$, 
P. Petit$^3$, H. Dai$^4$, A. Thess$^4$, R. E. Smalley$^4$,
A.R.M. Verschueren$^5$, S.J. Tans$^5$ and C. Dekker$^5$}

\address{$^1$Department of Physics,
University of Pennsylvania, Philadelphia, PA 19104-6272 \\
$^2$Department of Materials Science and Engineering and
Laboratory for Research on the Structure of Matter\\
University of Pennsylvania, Philadelphia, PA 19104-6272\\
$^3$Institut Charles Sadron, CNRS-ULP, 6 rue Boussingault,
67000 Strasbourg France\\
$^4$Center for Nanoscale Science and Technology, Rice Quantum Institute \\
Departments of Chemistry and Physics, Rice University, Houston TX 77251\\
$^5$Department of Applied Physics and DIMES,
Delft University of Technology\\
Lorentzweg 1, 2628 CJ Delft, The Netherlands }

\date{\today}
\maketitle

\begin{abstract}{ 
Nonchiral single wall carbon nanotubes with an ``armchair" wrapping 
are theoretically predicted to be conducting, and high purity samples consisting
predominantly of these tubes exhibit metallic behavior with an intrinsic
resistivity which increases approximately linearly with temperature over
a wide temperature range. Here we study the coupling of the conduction electrons 
to long wavelength torsional shape fluctuations, or twistons. A one dimensional
theory of the scattering of electrons by twistons is presented which predicts
an intrinsic resistivity proportional to the absolute temperature. Experimental
measurements of the temperature dependence of the resistivity are reported
and compared with the predictions of the twiston theory. 
}
\end{abstract}
 \pacs{85.42, 61.46, 36.40c, 73.20.D}
]

Since their discovery \cite{iijima}, there has been great interest in the
electronic properties of carbon nanotubes. These are nanometer scale
structures which can be understood as a single layer of graphite wrapped to form
 a cylindrical tube. This wrapping can be specified by two integers [M,N]
which define the superlattice translation corresponding to an elementary
 orbit around the waist of the cylinder.
Theory predicts that a single wall tube can exhibit insulating, semi-metallic
or metallic behavior depending on the choice of the integers M and N.  Using
 double laser ablation of Co- and Ni- doped graphite targets, a new catalytic
 route to the synthesis of these structures has been discovered  which now  allows
 the production of single-wall tubes \cite{thess}. This process produces bulk samples in which the [10,10]
 wrapping is predominant \cite{cowley}.
Tubes produced in this process self organize during deposition in
a two dimensional triangular close packed lattice forming  ropes (bundles of tubes), and ultimately
 mats (three dimensional samples of entangled ropes). The [10,10]
tubes are predicted to be metallic, and experimental evidence that
 unoriented bulk samples as well as individual ropes are metallic
 has been  presented in \cite{thess} and \cite{fischer}.
 Here we study the intrinsic intra-tubule scattering processes
 responsible for the resistivities of these systems and then present experimental measurements of
 the temperature dependence of the resistivity. We find
 that the coupling of the low energy electronic states to thermal \it shape fluctuations \rm
 of the tubes leads naturally to a resistivity which scales linearly with
 temperature even deep into the low temperature regime. This behavior is quite unconventional
 for a metal,  and is actually found 
 experimentally in tubule-derived samples down to a crossover temperature $\approx$ 10-100 K.
 The effects of inter-tube coupling on the electronic and vibrational
degrees of freedom responsible for this effect  are also briefly discussed below.

  Here we will focus on the $[N,N]$ ``armchair" tubes, which band theory predicts
 to be metallic \cite{metal,kanemele}. The low energy electronic structure of a single
armchair tube 
 consists of two pairs of one dimensional bands which cross the Fermi energy,
 and these can be described by the massless Dirac Hamiltonian:
\begin{equation}
H_e = \int dx \sum_{a,\sigma} i  \hbar v_F (
\psi_{a\sigma+}^\dagger \partial_x \psi_{a\sigma+} -
\psi_{a\sigma-}^\dagger \partial_x \psi_{a\sigma-} ).
\end{equation}
Here $\psi_{a\sigma\pm}$ describes a right (left) moving
electron with band index $a=1,2$ and spin $\sigma=\uparrow,\downarrow$.
$v_F$ is the Fermi velocity.

The electrical resistivity is determined by the dominant mechanism
for backscattering of electrons. In the presence of short range direct electron
electron interactions, the Hamiltonian (1) maps onto a two channel ``Hubbard ladder" 
\cite{noack,balentsfisher}.
The backscattering of electrons due
 to repulsive electron electron interactions has been studied within this model \cite{balentsfisher,krotov},
 and one finds that above
 a crossover temperature it leads to  a resistivity which scales linearly with temperature \cite{balentsfisher}.
Here we consider  a different and what we believe to be the dominant scattering process, namely
the coupling between electrons and elastic deformations of
the tubes.  Our theory  is the tubule analog to the Bloch Gruneisen (BG) theory of
 the  scattering of a Bloch electron
 by the low energy long wavelength acoustic modes of the lattice \cite{ziman} .  For the
tubules one finds that modes which twist the tube around its axis of symmetry
 locally compress and stretch bonds on the surface of the cylinder. 
  We have found 
that these elastic deformations couple the right moving and left moving electronic
 states of (1) \cite{kanemele} and are thus effective at backscattering
the electrons.  However, the dispersions of both the
 electronic and lattice degrees of freedom are unusual for these structures
 which, as we discuss below,  leads one naturally into a regime in which the modes responsible for the backscattering
 are always heavily thermally populated.  We find that  this implies a temperature
 dependent resistivity
 which is proportional to the absolute temperature even well below the nominal Debye temperature for this system, as observed experimentally, and in contrast to the  usual BG theory of a conductor. 

In a previous paper we have found that among the various possible low energy shape fluctuations
 of a tubule, only the twist provides an efficient backscattering mechanism for a
 conduction electron \cite{kanemele}. Thus we consider the scattering of electrons by thermally
excited long wavelength ``twistons", i.e. the acoustic torsional modes of the tubule.
The coupling between electrons and twist is given by \cite {kanemele}
\begin{equation}
H_{e-t} = \lambda \int dx
\sum_{a,\sigma} \nabla\theta  \
\left(\psi_{a\sigma+}^\dagger \psi_{a\sigma-}
+ {\rm h.c.}\right)
\end{equation}
where $\theta(x)$ is the angle of the twist at  a  position x along the tubule.
The coupling constant
for an $[N,N]$ tube is $\lambda = 3 N \beta \hbar v_F /4\pi$, where
$\beta = \partial \ln t / \partial \ln d$ describes the change
in the bond hopping amplitude $t$ with bond length $d$.  The
dynamics of long wavelength twistons may be described by the
continuum elastic Lagrangian,
\begin{equation}
L_t = {1 \over 2}\int dx \left[ M_t \dot\theta^2 - C_t (\nabla\theta)^2 \right],
\end{equation}
where $M_t$ is the moment of inertia per unit length of the tube and
$C_t$ is the twist modulus.  The twiston dispersion is then
$\omega_q = v_t q$ with $v_t = \sqrt{C_t/M_t}$.

The effect of twistons is rather unusual  because
they are the only long wavelength phonons which
couple the right and left moving electrons in the Dirac spectrum for this system \cite{kanemele}.
Unlike the phonon scattering in an ordinary  metal, which requires a phonon with
momentum $2 k_F$, backscattering from twistons is \it not \rm
quantum mechanically frozen out at low temperatures. 
Twiston scattering introduces a \it single \rm scattering
event which scatters an electron from the right to left moving branch,
as shown in the inset of Fig. 1. 
Moreover, since the momentum of a typical electron at temperature
$T$ is $k_B T/ v_F$, the energy of the relevant
twistons are of order $2 k_B T v_t / v_F$.
Since $v_t << v_F$, these phonons are always heavily thermally populated.
The system is thus effectively in the ``high temperature" limit for phonon scattering
even at temperatures well below the Debye temperature.

The backscattering rate for an electron with momentum $k$
may be computed from Fermi's golden rule to be
\begin{equation}
{1\over\tau} = 2\pi \lambda^2 \int {dq\over 2\pi}
         {v_t q\over C_t} \coth({\hbar \omega_q \over 2 k_B T})
         \delta \left(v_F(2k-q)\right)
\end{equation}
where we have ignored the small twiston energy $ \hbar v_t q$ in the delta
function.  This rate is independent of $k$ and linearly proportional
to $T$.  The one dimensional electrical resistivity is given by
$\rho_{1D} = (h/e^2)/(8 v_F \tau_{\rm tr})$, where for pure
backscattering the transport lifetime is
$\tau_{\rm tr} = \tau/2$.  We thus find
\begin{equation}
\rho_{1D} = {9\over{32\pi^2}} {h\over e^2}{\beta^2 \over { Nc_t}} k_B T,
\end{equation}
where $c_t=C_t/N^3$ is independent of $N$.

While the parameters in our theory have not yet been measured
for carbon nanotubes, they can be estimated using the corresponding
quantities known for graphite.  Based on the in plane shear
modulus of graphite we estimate $c_t = 18 {\rm eV\AA}$ \cite{dd}.
This predicts a velocity $ \hbar v_t = 0.09 {\rm eV\AA}$
which is equal to the speed of the in plane transverse acoustic phonon
of graphite.  In addition, we estimate $\hbar v_F = 5.3 {\rm eV\AA}$
and $\beta = 2.3 \cite{pietr}$.
For a rope of triangular close-packed [10,10] tubes with a lattice constant
17 $\rm \AA$, this leads to a temperature dependent
contribution to the three dimensional rope resistivity with slope
$d\rho_{3D}/dT = 0.005 \mu\Omega{\rm cm/K}$
 
Balents and Fisher have recently shown that Umklapp scattering
due to electron-electron interactions also leads to a resistivity which is linear
in temperature \cite{balentsfisher}.  Whereas the twiston resistivity scales as
$1/N$, the Umklapp resistivity is proportional to $1/N^2$.  Thus for sufficiently
large tubes the lattice effects should dominate.  Comparing the prefactors
we estimate that for $N=10$ the two have comparable magnitudes, with
the twiston resistivity larger by a factor of 4 \cite{scaling}.

The above discussion has focused on the resistivity of an isolated
tube.  This ignores three-dimensional effects for the dynamics 
of both the electrons and the phonons, which may be crucial for the 
correct interpretation of measurements on bundles (or ``ropes") of tubes.  
At low frequency, twistons on neighboring tubes should be  coupled elastically
which leads to a gap in the twiston spectrum as $q \rightarrow 0$.
This may be described within an Einstein model
for the inter-tube twistons, with the dispersion relation
$\omega_q = \sqrt{ v_t^2 q^2 + \omega_0^2}$.
The energy scale $\hbar \omega_0$ may be estimated by considering
the corresponding phenomena in graphite and in crystalline $C_{60}$.
In graphite the relevant zone boundary phonon has energy 4 meV \cite{dd},
whereas the energy of librons in crystalline $C_{60}$ span the range
2 - 6 meV \cite{prass}.

Coherent tunneling of electrons between the tubes also
destroys the nesting of the Fermi surface.
Given the bandwidth $W$, a phonon with wavevector as large as
$q \approx W/v_F$ is needed is needed to backscatter,
so that at low temperature, direct backscattering can ultimately
be frozen out over a large part of the Fermi surface. $W$ is difficult to estimate, because
it will depend on the details of the orientational
 registry between neighboring tubes.  However, it is unlikely that it will
be negligible for this system.  Solid phases of  $C_{60}$ have an interball electronic bandwidth of
order 0.5 eV \cite{erwin}.  In graphite tunneling between neighboring layers leads to
two interlayer bandwidths, one of order
1 eV and one with a  much narrower width of order 10 meV.
Electronic structure calculations within the local density approximation for  a three dimensional lattice of [12,12] tubes have estimated
a bandwidth of order 0.5 eV \cite {gonze}.  It is possible that because of the
 frustration of a five-fold symmetric [10,10] tube in a six-fold
coordinated environment, the intertube bandwidth may be slightly
 narrower.

Our estimates of the scattering rates, and thus
the  resistivity due to twiston scattering may be
generalized to include both of these effects.
In Fig. 1 we plot the resistivity as a function of temperature calculated using
our one-dimensional model, and recalculated including these three dimensional effects
for the representative parameters $\omega_0 =$ 4 meV and $W =$ .5 eV.  We find that
the resistivity  of the three dimensional system
is then essentially linear for $T>100 K$.  This is well below the
effective Debye temperature for the twistons which is
of order 1000 K. We find that in this system the one dimensional behavior can control 
the resistivity so long as $(v_t/v_F)  W > \hbar \omega_0$ as seems likely in
this system.
The  dynamics is then essentially one dimensional for $k_B T > \rm {max} \it (\hbar \omega_0 ,
(v_t/v_F) W)$.  We note that Umklapp scattering is suppressed for $k_B T<W$.  
Due to the small ratio $v_t/v_F \approx .02$, twiston scattering is more robust
in the presence of inter tube coherence.

\begin{figure}
\epsfxsize=3in
\centerline{\epsffile{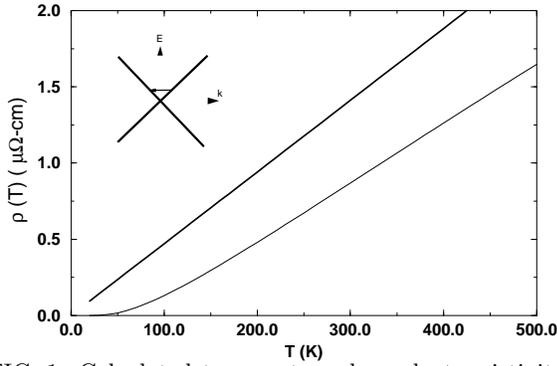}}
\caption{Calculated temperature dependent resistivity due to twiston scattering.
The upper bold curve is calculated for a one dimensional model for which $\rho\propto T$.
Including three dimensional intertube effects in both
 the electron and twiston degrees of freedom we obtain the lower curve.
which shows that the linear temperature dependence
 occurs in a three dimensional sample above relatively low crossover temperature.
The inset shows the process in which an electron scatters from
the right to left moving branch, emitting a long wavelength twiston.}
\end{figure}

To test the above theory, it is clearly desirable to measure the
electrical transport through a single isolated tube.  However, to date, single
tube transport has only been measured at very low temperature, where Coulomb
charging effects dominate\cite{tans}.  In Fig. 2 we present 4 different measurements
of the temperature dependence of the electrical resistivity of nanotube
ropes and unoriented bulk samples, all prepared as described in Ref. 2.
The top curve in Fig. 2(a) is a 4-probe 1 KHz 
measurement on a bulk sample using silver paint contacts.
Above about 200 K $\rho$ increases
linearly with temperature.  This confirms and extends to 580 K the
linear behavior previously observed up to 470 K \cite{fischer}.
The absolute value has little meaning since the tensor
components are not separable, the material is very porous, and
the effect of inter-rope contacts is unknown.  
The logarithmic derivative obtained from a linear fit in the interval
300 K $<$ T $<$ 580 K is 0.0008 K$^{-1}$.
The lower curve is derived from a microwave absorption measurement on 
a few micrograms of similar material.
The absolute value of $\rho$ from this technique depends on
the depolarization factor which is very difficult to estimate given the complex
morphology, while the logarithmic derivative, 0.001 K$^{-1}$ is comparable
to the 4-probe value. 

\begin{figure}
\epsfxsize=3in
\centerline{\epsffile{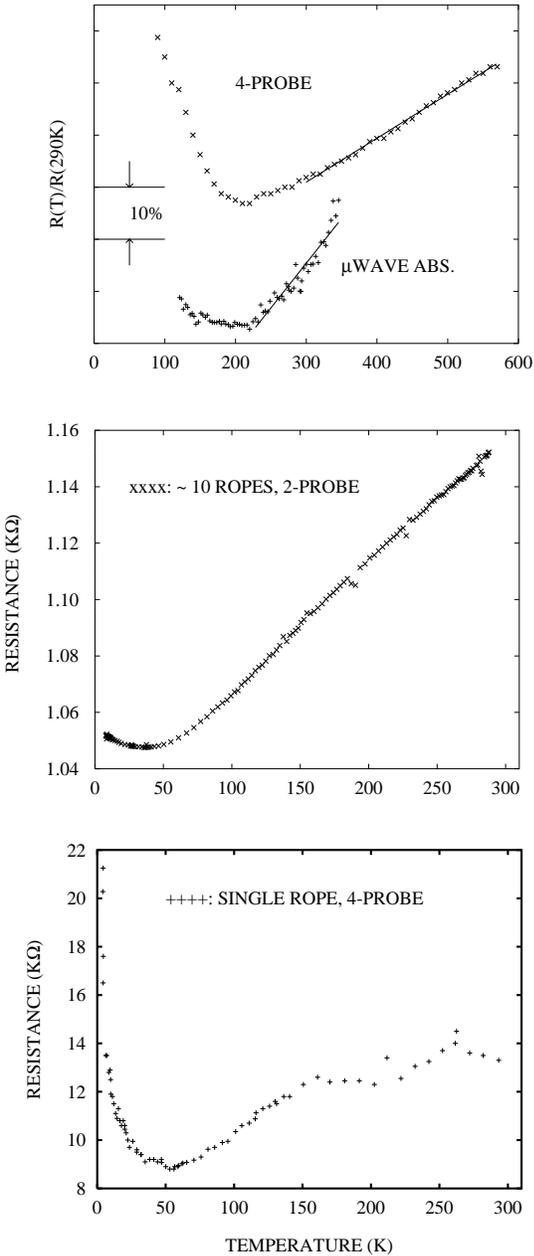}}
\vskip 2.0cm
\caption{Measured resistivities of samples of carbon nanotubes. 
(a) Bulk material:  The top curve is a 4-probe measurement.  
The lower curve is measured by microwave absorption.  
(b) 2-probe measurement on several ropes in parallel. 
(c) 4-terminal measurement of a single rope}
\end{figure}

Fig. 2(b) shows a 2-probe measurement of $\rho _{\parallel} $ (described
previously \cite{fischer}) on several ropes in parallel.  Again, linear
behavior is observed over a wide temperature range, with
a somewhat smaller logarithmic derivative, .0004 K$^{-1}$.  
4-probe absolute $\rho_{\parallel}$ 
measurements were performed at 300 K on similar samples
and span the range 30 - 100 $\mu\Omega $cm.  

Fig. 2(c) shows a 4-probe measurement 
on a single 7 nm diameter rope with voltage contacts 500 nm apart.  
The room temperature resistivity is $90 \mu\Omega{\rm cm}$,
which is consistent with the above measurement. 
The slope $d\rho/dT \approx .1 \mu\Omega{\rm cm/K}$.

These measurements clearly indicate metallic behavior
at high temperatures with $\rho$ increasing approximately
linearly with temperature.  Taking $\rho_{\parallel}$(300 K) = 
$90 \mu\Omega{\rm cm}$, and assuming the T dependence of the
bulk samples is dominated by $\rho_{\parallel}$(T),
we infer the absolute slope from the first three measurements.  
For the bulk, microwave and 2-probe rope measurements we thus find 
$d\rho/dT \sim .07, .09$ and $.04 \mu\Omega{\rm cm/K}$.
While the four measurements of the slope agree to within a factor
of 2.5, they are a factor of 8-20 larger than the twiston scattering 
theory prediction.  Part of this discrepancy could arise from
the presence of non-metallic tubes in the ropes.  
Recent electron diffraction measurements \cite{cowley} on similar materials
have indicated that more than 50\% of the tubes in a rope are chiral
and hence insulating \cite{metal,kanemele}.  
The presence of such ``dead'' tubes would lead to an overestimate of the rope's intrinsic
resistivity.   In addition, variations in the electron tunneling matrix
elements between different
tubes in a rope - which depend sensitively on the relative 
orientation of the tubes - could lead to an additional source 
of backscattering which is not present for a single tube.
In order to distinguish such effects from the intrinsic resistivity
of a single tube a high temperature transport measurement on a
single tube is clearly desirable.

It is striking that in addition to the high
temperature linear resistivity, all the experimental measurements show an upturn in the
resistivity at low temperature.  The onset of this low temperature behavior depends
on the sample morphology and can
be as low as 10K for
single ropes. It has been suggested that this upturn may signal
a condensation of the system to form a collective charge- or spin- density wave ground
state in the tube \cite{balentsfisher} \cite {krotov}. However, the observed dependence of this
crossover on the sample morphology and quality suggests that disorder or other three dimensional
effects may actually control this low temperature behavior. It will be important
to carry out further experimental work to 
understand the origin of this nonconducting low temperature behavior. 
 
    The work at the University of Pennsylvania was supported
under NSF Grant DMR 95-05425 (CLK) and by the Department of Energy under Grants
DE-FG02-84ER45118 (EJM) and DE-FC02-86ER45254 (JEF). The work at Rice was
supported by the ONR under Grant N0014-91-J1794 and by a grant from the Robert
A. Welch Foundation.  The work at Delft was supported by the Dutch Foundation for Fundamental
Research on Matter (FOM).

\end{document}